\documentclass[12pt]{article}
\usepackage{epsf}
\def\beq{\begin{equation}}
\def\eeq{\end{equation}}
\def\ap#1#2#3 {Ann. Phys. (NY) {\bf#1} (19#2) #3}
\def\err#1#2#3 {{\it Erratum} {\bf#1} (19#2) #3}
\def\ib#1#2#3 {{\it ibid.} {\bf#1} (19#2) #3}
\def\ijmp#1#2#3 {Int. J. Mod. Phys. {\bf#1} (19#2) #3}
\def\jetp#1#2#3 {JETP Lett. {\bf#1} (19#2) #3}
\def\mpl#1#2#3 {Mod. Phys. Lett. {\bf#1} (19#2) #3}
\def\np#1#2#3 {Nucl. Phys. {\bf#1} (19#2) #3}
\def\pl#1#2#3 {Phys. Lett. {\bf#1} (19#2) #3}
\def\prep#1#2#3 {Phys. Rep. {\bf#1} (19#2) #3}
\def\prev#1#2#3 {Phys. Rev. {\bf#1} (19#2) #3}
\def\prl#1#2#3 {Phys. Rev. Lett. {\bf#1} (19#2) #3}
\def\sjnp#1#2#3 {Sov. J. Nucl. Phys. {\bf#1} (19#2) #3}
\def\spj#1#2#3 {Sov. Phys. JETP {\bf#1} (19#2) #3}
\def\spu#1#2#3 {Sov. Phys. Usp. {\bf#1} (19#2) #3}
\def\zp#1#2#3 {Zeit. Phys. {\bf#1} (19#2) #3}

\begin{document}
\begin{titlepage}
\begin{center}
{\Large \bf Theoretical Physics Institute \\
University of Minnesota \\}  \end{center}
\vspace{0.2in}
\begin{flushright}
TPI-MINN-02/01-T \\
UMN-TH-2041-02 \\
February 2002 \\
\end{flushright}
\vspace{0.3in}
\begin{center}
{\Large \bf  Non-factorizable terms, heavy quark masses, and semileptonic decays of $D$ and $B$ mesons.
\\}
\vspace{0.2in}
{\bf M.B. Voloshin  \\ }
Theoretical Physics Institute, University of Minnesota, Minneapolis,
MN
55455 \\ and \\
Institute of Theoretical and Experimental Physics, Moscow, 117259
\\[0.2in]
\end{center}

\begin{abstract}
The non-factorizable terms in the operator product expansion have been recognized as one of theoretical obstacles for precision determination of the mixing parameter $V_{ub}$ from semileptonic $B$ decays. 
It is pointed out here that the recent CLEO data on the parameters of the heavy quark expansion $\lambda_1$ and ${\bar \Lambda}$, combined with a theoretical bound on $\lambda_1$ strongly favor the existence of a sizeable contribution of non-factorizable terms in semileptonic decays of $D$ mesons. Thus these terms are likely to solve the long-standing problem of the deficit of semileptonic decay rate of the $D$ mesons, and with better data their magnitude can be determined and used in studies of the parameter $V_{ub}$.
\end{abstract}

\end{titlepage}

The determination of the weak mixing parameter $V_{ub}$ from the data on inclusive semileptonic decays of $B$ mesons $B \to X_u \, \ell \, \nu$ requires kinematical cuts in order to discriminate against the dominant contribution of $B \to X_c \, \ell \, \nu$ (see e.g. in the review talks \cite{wise,ligeti}). The inclusive decay rate in the kinematically restricted part of the phase space however is sensitive to effects that being relatively small in the total rate are dominantly concentrated in the kinematical region of interest. A typical example of such behavior is presented by the method \cite{bll,bn} using a cut in the invariant mass of the lepton pair: $q^2 \ge (M_B-M_D)^2$, which leaves only about 20\% of the total inclusive decay rate of $B \to X_u \, \ell \, \nu$. When considered within the operator product expansion this inclusive decay rate however receives a contribution from a matrix element over the $B$ mesons of a non-factorizable four-quark operator. More specifically, this contribution reads as (see e.g. \cite{nu,mv} and references therein)
\beq
\delta \Gamma(B \to X_u \, \ell \, \nu) = 
{G_F^2 \, |V_{ub}|^2\, f_B^2 \, m_b^2 \, m_B \over 12 \, \pi} \, (B_2-B_1)~,
\label{nfb}
\eeq
where $f_B$ is the $B$ meson annihilation constant, and the phenomenological parameters ``bag constants" $B_1$ and $B_2$ parameterize the matrix elements of four quark operators over the $B$ meson\footnote{The nonrelativistic normalization: e.g. $\langle B | b^\dagger b | B \rangle=1$, for heavy quark operators is used throughout this paper.}:
\begin{eqnarray}
\left \langle B |({\bar b}\gamma_\mu (1-\gamma_5) u) ({\bar u}\gamma_\mu (1-\gamma_5) b) | B \right \rangle &=&{f_B^2 \, m_B \over 2} \, B_1~, \nonumber \\
\left \langle B |{\bar b} (1-\gamma_5) u) ({\bar u} (1+\gamma_5) b) | B \right \rangle &=& {f_B^2 \, m_B \over 2} \, B_2~.
\label{bagc}
\end{eqnarray}
In the factorization limit, where the matrix elements are saturated by the vacuum insertion between the two-quark factors, both constants $B$ are equal to one and cancel in the expression (\ref{nfb}). However the deviations from factorization are expected to be at a $\sim 10\%$ level. (E.g. these deviations are generally $O(N_C^{-2})$ in the limit of large number of colors $N_C$.)

At the expected level of about 10\% of breaking of the factorization relations, the natural magnitude of this contribution amounts to only about 2.5\% of the total rate, but it is all concentrated near the maximal value of $q^2$, formally $\delta(q^2-m_b^2)$, and thus is clearly greatly enhanced relative to the rate in the restricted part of the phase space. For this reason a better understanding of the magnitude of the non-factorizable terms is of a great importance for an accurate determination of $|V_{ub}|$ from the data.

The differences of the non-factorizable terms between heavy mesons with different light quark flavors, i.e. the flavor non-singlet part of these terms, can be tested \cite{mv} by measuring the differences of the semileptonic decay rates either between the $B$ mesons or between the $D$ mesons. An evaluation \cite{mv} of the non-singlet part of a similar non-factorizable term (although with a somewhat different color structure) from the difference of lifetimes of $D_s$ and $D_0$ mesons confirms that the deviations from factorization are indeed at a $\sim 0.1$ level.

An estimate of the light-flavor singlet part however requires an evaluation of the absolute decay rates of heavy mesons, rather than their differences, which also contain other, formally leading in the heavy quark expansion terms that are flavor-independent. Naturally, extracting the contribution of the non-factorizable terms from the absolute decay rates requires a quite precise knowledge of the parameters involved in the leading, perturbative term, as well as in the $O(m_Q^{-2})$ terms. This essentially leaves the semileptonic decays of $D$ mesons as the only appropriate testing ground for the flavor singlet part of the non-factorizable terms. 

The problem of explaining the total semileptonic decay rates of $D$ mesons has been discussed in the literature in the past \cite{vc,bds}, where it has been noticed that the theoretical decay rate, scaling with mass of the charmed quark essentially as $m_c^5$, is significantly lower than the experimental value if $m_c$ is taken to be around $1.4 \,  GeV$, as suggested by the QCD sum rules for charmonium\cite{six} and possibly by other considerations.  In order to remedy this contradiction Chernyak \cite{vc} suggested that the appropriate value of $m_c$ is substantially larger, $m_c \approx 1.65\, GeV$, while Blok, Dikeman and Shifman \cite{bds}, keeping $m_c$ not larger than about 1.4 GeV, considered and dismissed a possible contribution of non-factorizable terms, but rather attributed the contradiction to possible violation of the quark-hadron duality. Certainly a violation of the duality and a general failure of the OPE at a scale as low as the charmed quark mass is a logical possibility, in which case a consideration of the $B \to X_u \, \ell \, \nu$ decay rate with the cut $q^2 \ge (M_B-M_D)^2$ would also be of little meaning, since the scale involved in this case $\mu_c \approx m_c \, (1-m_c/2m_b)$ \cite{bn} is yet somewhat lower. However, since there are no other indications of such failure, it looks more constructive to still use the heavy quark expansion and to explore the possibility that it is the contribution of the non-factorizable terms that reconciles the experimental semileptonic widths of the $D$ mesons with the theoretical expression using a moderate value of $m_c$. The latter possibility, also discussed in Ref.\cite{nu}, can now be addressed with better certainty, due to recent CLEO data \cite{cleo} on the parameters $\lambda_1$ and $\bar \Lambda$ of the heavy quark expansion.

One can notice that when applied to semileptonic decay rates of the $D$ mesons, similarly to eq.(\ref{nfb}) the contribution of the non-factorizable terms can be written as
\beq
\delta \Gamma_{sl}(D )= {G_F^2 \,  f_D^2 \, m_c^2 \, m_D \over 12 \, \pi} \, \left ( |V_{cs}|^2 \, \delta B_s + |V_{cd}|^2 \, \delta B_d \right )~,
\label{nfc}
\eeq
where $\delta B_q$ is an analog of the difference $B_2 - B_1$ for  operators involving a light quark $q$, instead of the $u$ quark in eq.(\ref{bagc}). In the leading order in heavy quark mass the parameter $\delta B_q$ can be written in terms of the matrix element of an operator involving only spatial components (in the rest frame of the heavy meson) of the $V-A$ currents:
\beq
2 \left \langle P_Q | ({\bar Q}{\vec \gamma} (1-\gamma_5) q) ({\bar q}{\vec \gamma} (1-\gamma_5) Q) | P_Q \right \rangle =f_P^2 \, m_P  \, \delta B_q~,
\label{spat}
\eeq
with $P_Q$ standing for a pseudoscalar meson containing a heavy quark $Q$. Numerically, the expression (\ref{nfc}) gives
\beq
\delta \Gamma_{sl}(D)=0.08 \, ps^{-1} \, \left ({m_c \over 1.4 \, GeV} \right )^2 \, \left ( {f_D \over 0.2 \, GeV} \right )^2 \, \left ( {|V_{cs}|^2 \, \delta B_s + |V_{cd}|^2 \, \delta B_d \over 0.1} \right )~.
\label{nfcn}
\eeq
Thus  with  `natural' values $\delta B_s, \delta B_d \approx 0.1$ the effect of the non-factorizable terms easily reaches about one half of the experimental semileptonic decay rate, e.g. $\Gamma_{sl}(D^0)= 0.164 \pm 0.007 \, ps^{-1}$ \cite{pdg}. Therefore an analysis of these rates necessarily should include the non-factorizable terms even at their expected suppressed level.

As already mentioned the estimate of about 1.4 GeV for the `pole' mass of the charmed quark originates from the charmonium sum rules \cite{six}. Subsequent development of this method had however revealed that the perturbative expansion in $\alpha_s$ for the propagation of the $c \bar c$ pair, used in derivation of the sum rules, involves unusually large coefficients in the order $\alpha_s^2$ \cite{peter}. This might cast a general doubt on reliability of extraction of the parameter $m_c$ from the sum rules.

An alternative way of extracting the heavy quark mass parameter from the data is based on the mass formula for a heavy pseudoscalar meson:
\beq
M_P=m_Q+{\bar \Lambda} + {\mu_\pi^2 - \mu_g^2 \over 2 \, m_Q}+O(m_Q^{-2})~,
\label{mf}
\eeq
where ${\bar \Lambda}$ is the mass shift due to light degrees of freedom in the meson, and $\mu_\pi^2$ and $\mu_g^2$ describe the kinetic and chromomagnetic energy of the heavy quark inside the meson: $\mu_\pi^2 = -\langle P | ( {\bar Q} {\vec D}^2 Q)|P \rangle$, $\mu_g^2 = \langle P | ( {\bar Q} ({\vec \sigma} \cdot {\vec B})  Q)|P \rangle$ with $D_\mu$ being the QCD covariant derivative and ${\vec B}$ being the chromagnetic field operator (see e.g. in the review \cite{bsu}). The parameter $\mu_g^2$ is readily found from the mass splitting of vector and pseudoscalar mesons, $\mu_g^2 \approx 0.37 GeV^2$, while the values of the parameters ${\bar \Lambda}$ and $\mu_\pi^2$ are less certain. 

The value of $\mu_\pi^2$ is however constrained by the inequality \cite{mv2}
\beq
\mu_\pi^2 \ge \mu_g^2~,
\label{ineq}
\eeq
which follows, in terms of the nonrelativistic expansion for a heavy quark, from the non-negativity of the Pauli Hamiltonian $-({\vec \sigma} \cdot {\vec D})^2 = -{\vec D}^2 - ({\vec \sigma} \cdot {\vec B})$, or from a more general inequality, not necessarily related to the heavy quark expansion,
\beq
{\partial M_P \over \partial m_Q}= \langle P | ({\bar Q} Q) |P \rangle \le \langle P | (Q^\dagger Q) |P \rangle =1~.
\label{ineq2}
\eeq
When applied to the expansion (\ref{mf}) the latter general inequality reduces to the constraint (\ref{ineq})\footnote{It can be noticed that the general inequality (\ref{ineq2}) implies the bound \cite{mv3} on the difference of the $b$ and $c$ quark masses: $m_b-m_c \ge M_B - M_D \approx 3.41\, GeV$, which does not rely on the heavy quark expansion.}.

Recently CLEO has published \cite{cleo} an analysis of the parameters ${\bar \Lambda}$ and $\lambda_1 = -\mu_\pi^2$ from their data on the inclusive photon spectrum in the decays $B \to X_s \, \gamma$ and on the second moment of the hadronic invariant mass in the semileptonic decays of $B$ mesons\footnote{It should be noted, for the sake of rigor, that depending on definition of perturbative renormalization contributions there may arise a difference between $\lambda_1$ and $-\mu_\pi^2$, as discussed in Ref.\cite{bsu}. However in the context of the analysis of Ref.\cite{cleo}, no such difference arises, and $\lambda_1$ and $-\mu_\pi^2$ are equivalent. We switch to the notation $\lambda_1$ in the rest of this paper in order to facilitate following the relation to Ref.\cite{cleo}.}. The results of this analysis can be converted, using the mass formula (\ref{mf}), into a value of $m_c$, which in turn can be used in the expression for the total semileptonic decay rate of the $D$ mesons and thus enables an evaluation of the contribution of the non-factorizable terms from comparison with the experimental rate. A calculation along these lines is to be presented in the rest of this paper. The additional element in this analysis as compared to Ref.\cite{cleo} is that the constraint (\ref{ineq}), $\lambda_1 \le -0.37$, is explicitly taken into account. 

The expression for the semileptonic decay rate of a $D$ meson, including the perturbative QCD corrections up to two loops \cite{rit} and the terms in heavy quark expansion of order $m_c^{-2}$ \cite{buv} as well as the non-factorizable terms of order $m_c^{-3}$ can be written as
\begin{eqnarray}
&&\Gamma_{sl}(D)= {G_F^2 \, m_c^5 \over 192 \, \pi^3} \left [ |V_{cs}|^2 \, \left (1-8 {m_s^2 \over m_c^2} \right ) + |V_{cd}|^2 \right ] \times
\nonumber \\ 
&&\left [ 1 - 2.413 \, {\alpha_s \over \pi} - 23.44 \, \left ({\alpha_s \over \pi} \right )^2 \right ] \, \left (1+ {\lambda_1+\mu_g^2 \over 2 \, m_c^2} \right ) \, \left ( 1 - {\mu_g^2 \over 2 \, m_c^2} \right ) + \delta \Gamma_{sl}(D)~,
\label{gsl}
\end{eqnarray}
where $\alpha_s=\alpha_s(m_c)$,  $\delta \Gamma_{sl}(D)$ is given by eq.(\ref{nfc}), and a certain inaccuracy has to be admitted in the treatment of the cross terms between e.g. the radiative corrections and the effect of the finite mass $m_s$ of the strange quark or between the radiative corrections and a part of the $O(m_c^{-2})$ terms. This inaccuracy however is at the level of other uncertainties involved in eq.(\ref{gsl}), e.g. due to higher perturbative terms, or the experimental uncertainties in the data, and can be safely neglected in the present calculation.
\begin{figure}[ht]
  \begin{center}
    \leavevmode
     \epsfxsize=3.5in
     \epsfbox{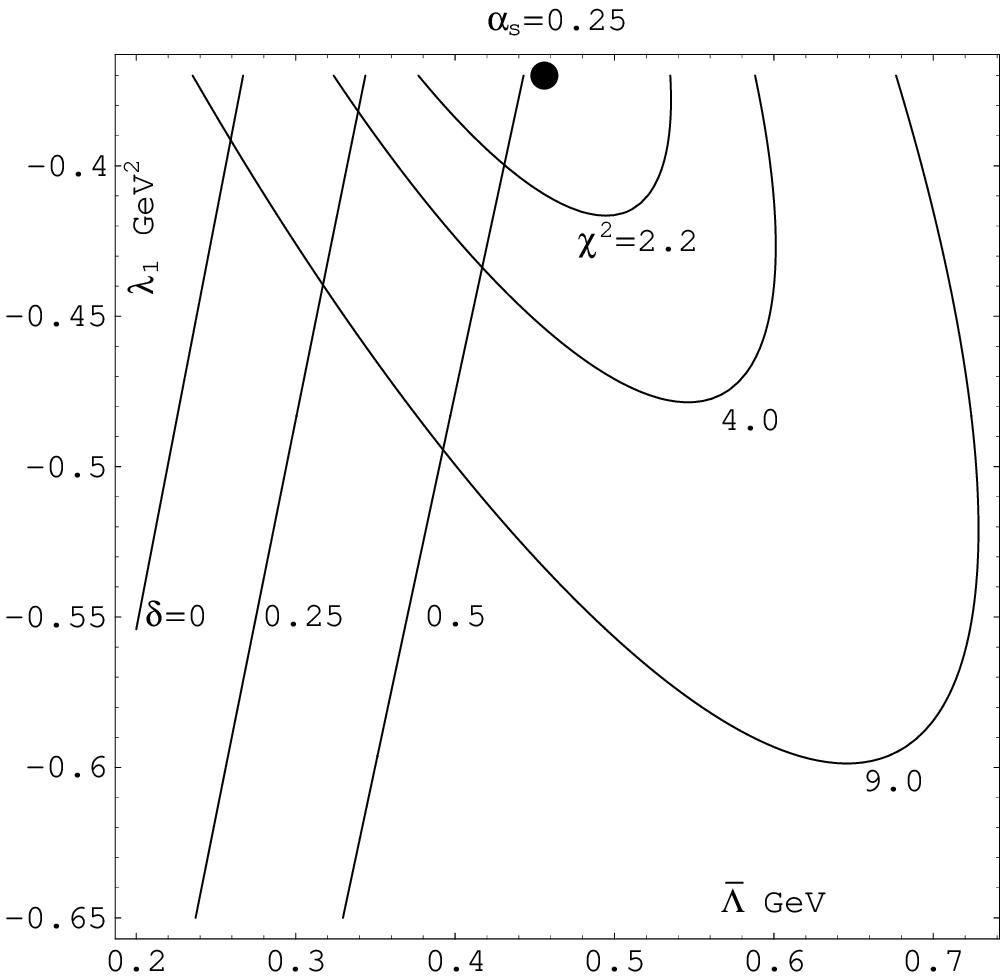}
     \epsfxsize=3.5in
     \epsfbox{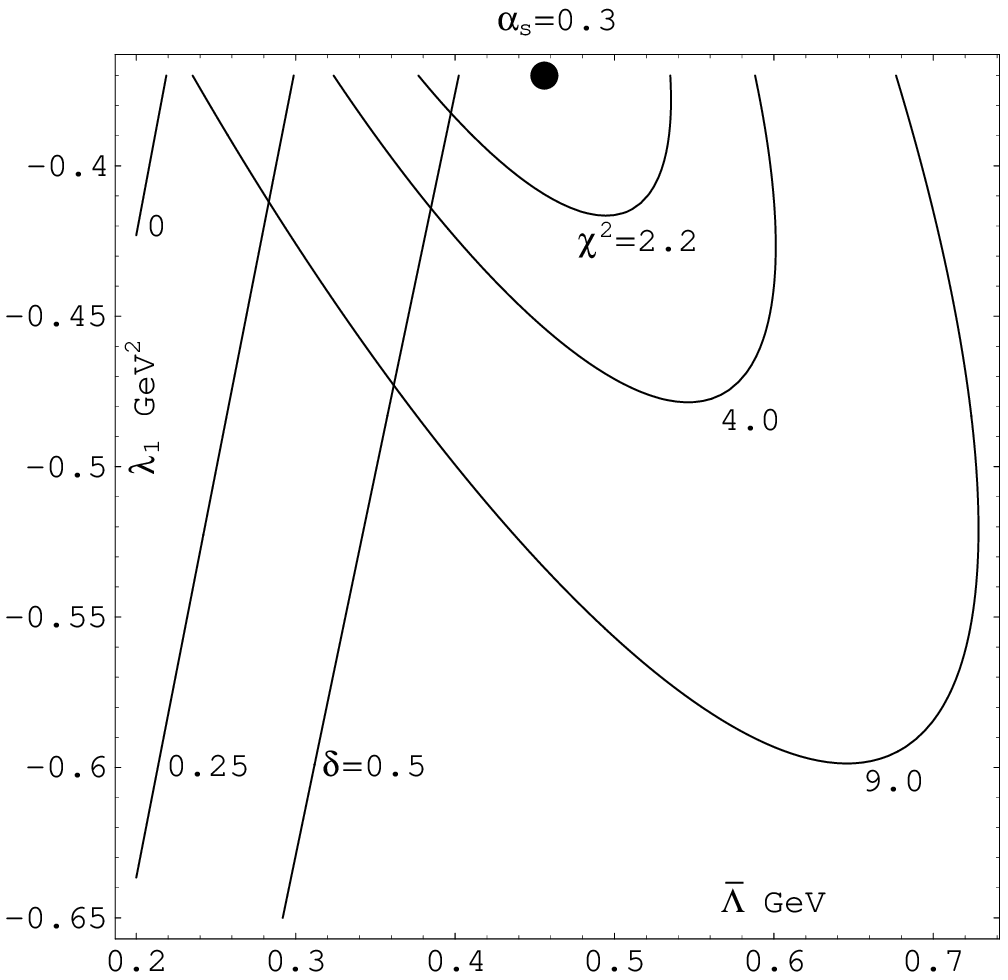}
    \caption{The lines of constant relative contribution $\delta$ of non-factorizable terms to $\Gamma_{sl}(D)$ overlayed on the CLEO results  \cite{cleo} in the physical region $\lambda_1 \le -0.37$ at $\alpha_s(m_c)=0.25$ and $\alpha_s(m_c)=0.3$. The elliptical arcs correspond to the indicated levels of $\chi^2$ in the CLEO data. The filled circle is centered at the minimum of $\chi^2$ of about 1.2 in the physical region. The lines in the left part of the plots correspond to the indicated values of $\delta$.}
    \label{fig:plots}
  \end{center}
\end{figure} 

The equation (\ref{gsl}) can be used for evaluating the contribution of the non-factorizable terms as a function of $\lambda_1$ and ${\bar \Lambda}$, since these also determine the mass $m_c$ through the mass formula (\ref{mf}), and the actual rate $\Gamma_{sl}(D)$ is known experimentally. The results of such evaluation are shown together with the CLEO data \cite{cleo} in Fig.1 in terms of the dimensionless ratio $\delta=\delta \Gamma_{sl}(D)/\Gamma_{sl}(D^0)$, where the experimental value is used for $\Gamma_{sl}(D^0)$ in the denominator. Only the physical region $\lambda_1 \le -0.37$ is shown in the plots. Although the central point of the CLEO result lies above this region, the minimal value of $\chi^2$ in the physical region is about $1.2$, so that the data are well consistent with the constraint (\ref{ineq}). The parameters corresponding to this point of maximum likelihood are ${\bar \Lambda} =0.456 \, GeV$ and (naturally) $\lambda_1=-0.37$, saturating the bound (\ref{ineq}). According to the mass formula (\ref{mf}) these values translate into $m_c \approx 1.41 \, GeV$ and, when used in eq.(\ref{gsl}) for the $\Gamma_{sl}(D)$, the maximum probability corresponds to $\delta \approx 0.5$ if $\alpha_s(m_c)=0.25$ and to $\delta \approx 0.6$ if $\alpha_s(m_c)=0.3$. (Clearly, the estimated value of $\delta$ grows with $\alpha_s(m_c)$, since the QCD radiative corrections suppress the perturbative and $O(m_c^{-2})$ terms in the rate.) It can be also noted that the point of maximum likelihood corresponds to the `best value' of the appropriate `pole' mass of the $b$ quark $m_b \approx 4.82 \, GeV$, in a reasonable agreement with the evaluation from the QCD sum rules for $\Upsilon$ resonances \cite{mv4}.

The most likely values of $m_c$ and $\delta$ suggested by the CLEO data are in a very good agreement with theoretical expectations from other considerations. In particular the value of $\delta$ around $0.5-0.6$ agrees with the expectation that the suppression of the non-factorizable terms is about 0.1.
As seen from the plots in Fig.1 the value $\delta=0$, and hence the mass $m_c$ around $1.6-1.65 \, GeV$, has quite low likelihood and is only marginally consistent, at a level of about $3\sigma$, with the CLEO data and the constraint (\ref{ineq}).
Clearly, a more statistically significant evaluation of the parameters of the heavy quark expansion and thus of the quantity $\delta$ requires improvement of both the theoretical and experimental accuracy in an analysis along the lines of that in Ref.\cite{cleo} supplemented by the remarks presented in this paper. In any event, a precision determination of the weak mixing parameter $|V_{ub}|$ from a constrained spectrum of semileptonic $B$ decays is quite unlikely before the issues related to semileptonic decay rates of $D$ mesons are resolved both theoretically and experimentally.

I thank A. Czarnecki for an update on the status of perturbative calculations of semileptonic decay rates, R. Poling for a discussion of the CLEO work \cite{cleo} and D. Cronin-Hennessy for e-mail communication concerning the results of the CLEO analysis and the parameters of the final plot in their paper. This work is supported in part by DOE under the grant
number DE-FG02-94ER40823.


\end{document}